# Superconductivity in single-layer films of FeSe with a transition temperature above 100 K

Jian-Feng Ge[1], Zhi-Long Liu[1], Canhua Liu[1]*, Chun-Lei Gao[1], Dong Qian[1], Qi-Kun Xue[2]*, Ying Liu[1,3], Jin-Feng Jia[1]*

[1]Key Laboratory of Artificial Structures and Quantum Control (Ministry of Education), Department of Physics and Astronomy, Shanghai Jiao Tong University, 800 Dongchuan Road, Shanghai 200240, China.

[2]Department of Physics, Tsinghua University, Beijing 100084, China.

[3]Department of Physics and Materials Research Institute, Pennsylvania State University, University Park, PA 16802, USA.

*Correspondence to: canhualiu@sjtu.edu.cn; qkxue@mail.tsinghua.edu.cn; jfjia@sjtu.edu.cn.

**Recently, interface has been employed to enhance superconductivity in the single-layer FeSe films grown on SrTiO$_3$(001) (STO) with a possible $T_c$ of ~ 80 K (ref. 1) , which is nearly ten times of the $T_c$ of bulk FeSe and is above the $T_c$ record of 56 K for the bulk Fe-based superconductors[2]. This work together with those on superconducting oxides interfaces[3-6] revives the long-standing idea that electron pairing at a two-dimensional (2D) interface between two different materials is a potential path to high transition temperature ($T_c$) superconductivity[7-12]. Subsequent angle-resolved photoemission spectroscopy (ARPES) measurements revealed different electronic structure from those of bulk FeSe[13-15] with a superconducting-like energy gap closing at around 65K. However, previous *ex situ* electrical transport measurements[1, 15] could only detect the zero-resistance below ~30 K. Here we report the observation of high $T_c$ superconductivity in the FeSe/STO system. By *in situ* 4-point probe (4PP) electrical transport measurement that can be conducted at an arbitrary position of the FeSe film on STO, we detected superconductivity above 100 K. Our finding makes FeSe/STO the exciting and ideal research platform for higher $T_c$ superconductivity.**

Search for superconductors with a $T_c$ above the liquid nitrogen temperature (77 K) led to the discovery of a high-$T_c$ cuprate with a $T_c$ above 130 K over two decades ago[16]. Even though the value of $T_c$ is only 26 K in the first Fe-based superconductor, LaFeAsO (ref. 17), the subsequent work on a series of Fe-based superconductors showed the highest $T_c$ of the Fe-based superconductors under ambient pressure found in SmFeAsO is as high as 56 K (ref. 2). So far, superconductors with a $T_c$ > 77 K have been limited to cuprates. Recently, interface effects were employed to enhance the superconductivity in FeSe. Single-layer films of FeSe grown on a

SrTiO$_3$(001) substrate, referred to below as FeSe/STO, were found to exhibit a superconducting energy gap, Δ, as large as 20.1 meV, which was detected by *in situ* scanning tunneling microscopy/spectroscopy (STM/STS) measurements at 4.2 K (ref. 1). A $T_c$ value as high as 86 K would be expected if the ratio of $2\Delta/k_BT_c = 5.5$, where k$_B$ is the Boltzmann constant, found in bulk FeSe [with a $T_c$ of 9.4 K (ref. 18)] were applicable for the FeSe/STO. This work, together with the earlier work on superconducting oxides interfaces[3-6], demonstrates that interface between two different materials provide not only a rich system for studying two-dimensional (2D) superconductivity, but also a potential pathway to high-$T_c$ superconductivity[7-12].

Indeed, recent angle-resolved photoemission spectroscopy (ARPES) experiments on the FeSe/STO system revealed different electronic structure from those of bulk FeSe and possible occurrence of superconductivity around 65 K (refs. 13,14). An *ex situ* transport measurements performed on FeSe/STO protected by multiple layers of FeTe and amorphous Si overlay revealed a zero-resistance $T_c$ of 23.5 K and an onset $T_c > 40$ K (ref. 15). Evidently the addition of protection layers suppresses superconductivity in single-layer FeSe. In this work, we report electrical transport measurements on single-layer films of FeSe grown on Nb-doped SrTiO$_3$ substrate using an *in situ* 4-point probe (4PP) technique. We found that superconductivity could be obtained even at a temperature as high as 109 K.

Single-layer films of FeSe were grown on Nb-doped SrTiO$_3$(001) surface by the same method as reported previously[1], employing extra Se flux in an MBE system equipped with STM/STS and 4PP capabilities. The growth process was monitored by reflection high-energy electron diffraction (RHEED) (Fig. 1a), which allows the precise control of film growth needed to achieve one unit-cell thickness as exactly as possible. The crystal nature of the films was confirmed by STM imaging at both large and atomic scales, as shown in Figs. 1b and 1c, respectively.

The 4PP technique has been used to study the superconducting single-layer Pb films grown on Si(111) surfaces[19]. There are two basic measurement configurations that have been proposed ever since 1950's (refs. 23, 24, Supplementary Information): C1423 and C1234 (Fig. 2a). In C1423 (C1234), a D.C. current, $I_{14}$ ($I_{12}$), is applied through Tips 1 and 4 (1 and 2) while voltage drop, $V_{23}$ ($V_{34}$), is measured between Tips 2 and 3 (3 and 4). While the applicability of the 4PP technique to extremely thin films was well documented[12], the result of the 4PP measurements may depend on the conditions of the contacts. In C1423 configuration, for example, the two voltage tips 2 and 3 register a finite voltage difference (Fig. S1h) even in the case that the film is superconducting. This may occur if the mechanical motion of the tips making contact with the film is too strong to break the FeSe film. Therefore, when 4PP detects a zero voltage on a film grown on a conducting substrate, it does mean zero resistance of the film as the film shorts the conducting substrate. However, when the 4PP detects a finite voltage, it may not necessarily mean that the sample is not superconducting. Indeed, the 4PP technique is a powerful tool for investigating superconductivity in films that cannot be taken out of a UHV system or an interface that is not accessible by surface probes.

Two typical I-V curves collected at 3 K in C1423 and C1234 are shown in Figs. 2b and 2c, respectively. The data demonstrate explicitly that the film is superconducting, with the critical current ($I_c$) defined by the current value for which the superconducting top layer can no longer short the conducting substrate with a finite resistance. Even though interpretation of the finite voltage in the superconducting I-V curves are complicated, the essentially zero voltage seen at low currents cannot be resulted from artificial effects of the contact, since all four tips of the 4PP have Ohmic contacts with the sample individually (Supplementary Information). It is also interesting to note that the extracted $I_c$ have similar values for both measurement configurations, which may be due to the conducting substrates suppress the difference between the two configurations. We performed the same 4PP measurements on an optimally doped $Bi_2Sr_2CaCu_2O_{8+\delta}$ single crystal ($T_c$ ~91 K), and obtained similar $I_c$ values for two different configurations as well (Supplementary Information). Moreover, the similar shape of I-V curves and the same $T_c$ as reported strongly indicate our measurements are reliable.

In addition, linear I-V curves (see Figs. S1 and S2) were obtained occasionally even at very low temperatures, possibly due to the inhomogeneity of the sample or due to hard contacts between a tip or tips and the sample, which results in local damages in the FeSe single layer. We have confirmed the latter case by intentionally pressing further the 4PP until the I-V curves change from a superconducting one to a linear one, as shown in Fig. S2. In the present experiment, we move the 4PP to various locations to collect I-V curves at a fixed temperature. The film was designated superconducting when a superconducting I-V curve was obtained at one of the locations. We found that if one superconducting I-V curve could be obtained in one location, it can be obtained in some other locations.

The results of the electrical transport measurements on the FeSe/STO are summarized in Fig. 3. At temperatures above 109 K, the I-V curves are linear in the full measured current range, in strong contrast to those of the superconducting ones taken below 109 K, as Fig. 3a shows. Therefore, we can determine the $T_c$ of the single-layer FeSe to be ~109 K. To make connection of the more traditional resistance vs. temperature plot, we also plot the resistance values extracted from the linear fit to I-V curves taken at fixed temperatures (Fig. 3b). To complete the R-T plot shown in Fig. 3b, the I-V curves were collected at various locations to avoid local damages on the FeSe film due to the possible tip-sample mechanical shift while the temperature was changing. We did the 4PP measurement on another sample with the tips fixed on one location while collecting the I-V curves at fixed temperatures for the R-T plot, in which more data were obtained and the $T_c$ is around 99 K (Fig. S4). It should be emphasized, however, that the value of the conduction in FeSe/STO above 109 K is dominated by that of the STO substrate rather than the FeSe film as the conducting substrate has much larger cross section area. Essentially, below $T_c$, the sample is shorted by the superconducting film of FeSe while above $T_c$, the sample is dominated by the conducting substrate. In fact this is the reason for the apparent sharp transition around 109 K in Fig. 3b. Indeed, while the 4PP does reveal a superconducting phase transition in the single-layer FeSe film, the normal-state resistance cannot be determined using this technique if the film is grown on a conducting substrate.

A cubic-tetragonal phase transition was observed previously on the vacuum annealed surface of STO at 105 K (ref. 20), raising the question on impact of this transition in the substrate. While the effect of the structural transition on the superconductivity in single-layer films of FeSe is yet to be understood, our extensive 4PP control measurements on bare STO (Insets of Figs. 3a and 3b) reveal no superconducting I-V curves, ruling out the possibility that the superconducting I-V curves observed on single-layer FeSe grown on STO actually came from the substrate. One may notice that the normal-state resistance extracted from the I-V curves shown in Figs. 2b and 2c are ~5 times more than that in Fig. 3a in the normal state. This is because they were collected using two 4PPs with different probe separations: 10 μm and 100 μm, respectively. Essentially the 4PP with a shorter tip separation is more sensitive to the film than that with a larger separation, yielding a larger normal-state resistance since the FeSe single-layer film in a normal state is less conducting than the bulk conducting STO substrate.

We emphasize that, below 109 K, we could always obtain superconducting I-V curves, in which the resistance keeps at zero while $I_c$ increases gradually from 0.2 mA at 109 K to 4.0 mA at 3 K, as shown in Fig. 3c. The temperature dependence of $I_c$ can fit to the Ginzburg-Landau model, $I_c(T)=I_c(0)(1-T/T_c)^{3/2}$ (ref. 21), which gives $I_c(0)=3.8 \pm 0.2$ mA and $T_c=111 \pm 4$ K. This agrees well to the $T_c$ value obtained directly from the I-V curves obtained at fixed temperatures. Furthermore, we investigated the superconducting I-V curves under an external magnetic field. It was found that $I_c$ decreased with increasing magnetic field, which is seen to remain nonzero up to 11 T (Fig. 3d). As shown in Fig. 3e, a superconducting I-V curve can still be observed at 11 T at 3 K, demonstrating the robustness of superconductivity in single-layer FeSe. This indicates a large out-of-plane $H_{c2}$ for this system, consistent with results obtained in previous *ex situ* transport measurements[18].

In summary, our *in situ* 4PP electrical transport measurements showed that the single-layer films of FeSe grown on a conductive STO substrate is superconducting at a temperature as high as 109 K. While the mechanism of the high-$T_c$ superconductivity in single-layer FeSe/STO is yet to be understood, interface enhanced electron-phonon coupling may have played a role in this remarkable phenomenon[1,7-9,22] giving a strong boost to the search of high-$T_c$ superconductivity in artificial interface systems and the development of high-temperature superconducting electronic applications in this relatively simple material system.

**Methods**

This experiment was carried out in an STM-MBE system (made by Unisoku Co.) consisting of two connected ultrahigh vacuum (UHV) chambers with a base pressure of $3\times10^{-10}$ Torr. A self-designed 4PP can be mounted onto the STM tip stage for *in situ* electrical transport measurement while the original STM working function is retained. Single-layer FeSe films were grown on the $TiO_2$ terminated and Nb-doped $SrTiO_3$ (001) (STO) substrate following the method reported previously[1]. A single-crystal STO chip was ultrasonically cleaned with ethanol and acetone first. After over-night degassing at 600°C in UHV, the STO wafer was kept at 950°C under Se atmosphere for 30 minutes and then gradually cooled down to 550°C for FeSe growth. RHEED was used to monitor the growth process, during which Se and Fe were co-deposited onto the STO surface with a flux rate of ~20:1. A post-annealing at the growth temperature was carried out for 30 minutes after the Se and Fe fluxes were stopped. The film was immediately transferred from the MBE growth chamber to the STM chamber for STM and 4PP transport measurements. A chemically etched tungsten tip was used for the STM and STS experiments.

The 4PP composed of 4 tips are in turn linearly aligned in a roughly equal separation. Two 4PPs were used in the experiments. One is made of Cu wire with a tip separation of ~100 μm, and the other is made of silicon cantilevers coated by Au with a tip separation of ~10 μm. These 4 tips are fixed on a ceramic plate mounted on a 4PP holder that we designed to contain four electrodes. We redesigned the tip stage of the STM system so that the 4PP tip holder can also be inserted onto the STM tip stage *in situ* without affecting the STM function when the STM tip holder is inserted back to the stage. The approach of the 4 tips toward the sample surface was precisely controlled by using the STM approaching system. The four tips are inclined to the sample surface with an angle of ~20º so that they can touch the sample surface gently even though the four tips are not of exactly the same length. In the transport measurement, I-V curves were collected at a fast rate (about 20 s per curve) while the temperature of the system was raised slowly (around 3 K/hour). In the 4PP electrical measurement, the electric current was sourced by Keithley 2400 SouceMeter while the voltage was measured by Keithley 2182A Nanovoltmeter. The electronics are capable of detecting sample resistance on the order of $10^{-6}$ Ω. To eliminate the thermal voltages in the circuit, the DC measurement current was reversed in the I-V curve measurements.

The thermometer used in the experiment is the Cernox Resistor type provided by Lake Shore Cryotronics (Model: CX-1030-SD-HT-0.3M). During this experiment, the thermometer reads 295.7 K at room temperature and 77.6 K when immersed in the liquid nitrogen. Given that the thermometer is thermally anchored at the sample stage that is thermally linked to liquid $^4$He or $N_2$ bath and the film sits on top of the substrate, the sample temperature cannot be lower than the cold stage temperature measured by the thermometer.

**References**

1. Wang, Q. Y. *et al.* Interface-induced high-temperature superconductivity in single unit-cell FeSe films on SrTiO$_3$. *Chin. Phys. Lett.* **29**, 037402 (2012).

2. Wu, G. *et al.* Superconductivity at 56 K in samarium-doped SrFeAsF. *J. Phys.: Condens. Matter* **21**, 142203 (2009).

3. Reyren, N. *et al.* Superconducting interfaces between insulating oxides. *Science* **317**, 1196-1199 (2007).

4. Gozar, A. *et al.* High-temperature interface superconductivity between metallic and insulating copper oxides. *Nature* **455**, 782-785 (2008).

5. Kozuka, Y. *et al.* Two-dimensional normal-state quantum oscillations in a superconducting heterostructure. *Nature* **462**, 487-490 (2009).

6. Richter, C. *et al.* Interface superconductor with gap behaviour like a high-temperature superconductor. *Nature* **502**, 528-531 (2013).

7. Ginzburg, V. L. On surface superconductivity. *Phys. Lett.* **13**, 101-102 (1964).

8. Cohen, M. H. & Douglass, D. H. Superconductive pairing across electron barriers. *Phys. Rev. Lett.* **19**, 118 (1967).

9. Strongin, M. *et al.* Enhanced superconductivity in layered metallic films. *Phys. Rev. Lett.* **21**, 1320 (1968).

10. Zhang, T. *et al.* Superconductivity in one-atomic-layer metal films grown on Si(111). *Nature Phys.* **6**, 104-108 (2010).

11. Uchihashi, T., Mishra, P., Aono, M. & Nakayama, T. Macroscopic superconducting current through a silicon surface reconstruction with indium adatoms: Si(111)-($\sqrt{7}\times\sqrt{3}$)-In. *Phys. Rev. Lett.* **107**, 207001 (2011).

12. Yamada, M., Hirahara, T. & Hasegawa, S. Magnetoresistance measurements of a superconducting surface state of In-induced and Pb-induced structures on Si(111) *Phys. Rev. Lett.* **110**, 237001 (2013).

13. He, S. L. *et al.* Phase diagram and electronic indication of high-temperature superconductivity at 65 K in single-layer FeSe films. *Nat. Mater.* **12**, 605-610 (2013).

14. Tan, S. Y. *et al.* Interface-induced superconductivity and strain-dependent spin density waves in FeSe/SrTiO$_3$ thin films. *Nat. Mater.* **12**, 634-640 (2013).

15. Zhang, W. *et al.* Direct observation of high-temperature superconductivity in one-unit-cell FeSe films. *Chin. Phys. Lett.* **31**, 017401 (2014).

16. Schilling, A., Cantoni, M., Guo, J. D. & Ott, H. R. Superconductivity above 130 K in the Hg–Ba–Ca–Cu–O system. *Nature* **363**, 56-58 (1993).

17. Kamihara, Y., Watanabe, T., Hirano, M. & Hosono, H. Iron-based layered superconductor La[O$_{1-x}$F$_x$]FeAs (x = 0.05−0.12) with $T_c$ = 26 K. *J. Am. Chem. Soc.* **130**, 3296-3297 (2008).

18. Song, Y. J. *et al.* Superconducting properties of a stoichiometric FeSe compound and two anomalous features in the normal state. *J. Korean Phys. Soc.* **59**, 312-316 (2011).

19. Hasegawa, S. *et al.* Electrical conduction through surface superstructures measured by microscopic four-point probe. *Surf. Rev. Lett.* **10**, 963-980 (2003).

20. Nestler, T. *et al.* Increased cubic–tetragonal phase transition temperature and resistivity hysteresis of surface vacuum annealed SrTiO$_3$. *Appl. Phys. A* **105**, 103 (2011).

21. Tinkham, M. *Introduction to Superconductivity* Ch. 4 (Dover Publications, New York, 2nd ed., 2004).

22. Lee, J. J. *et al.* Evidence for pairing enhancement in single unit cell FeSe on SrTiO$_3$ due to cross-interfacial electron-phonon coupling. Preprint at http://arxiv.org/abs/1312.2633 (2013).

23. Smits, F. M. Measurement of sheet resistivities with the four-point probe. *Bell Syst. Tech. J.* **37**, 711-718 (1958).

24. Valdes, L. B. Resistivity measurements on germanium for transistors. *Proc. I. R. E.* **42**, 420-427 (1954).

**Supplementary Information** is available in the online version of the paper.

**Acknowledgments:** We acknowledge discussions with Xuchun Ma, Lili Wang, Wenhao Zhang, Donglai Feng, and Tony Leggett. Financial support from the National Basic Research Program of China (Grants No. 2012CB927401, No. 2011CB921902, No. 2013CB921902, No. 2011CB922200, and No. 2012CB927403), NSFC (Grants No. 11374206, No. 91021002, No. 11274228, No. 10904090, No. 11174199, and No. 11134008, and No. 11274229), the Shanghai Committee of Science and Technology, China (Grants No. 12JC1405300, No. 13QH1401500, and No. 10JC1407100).

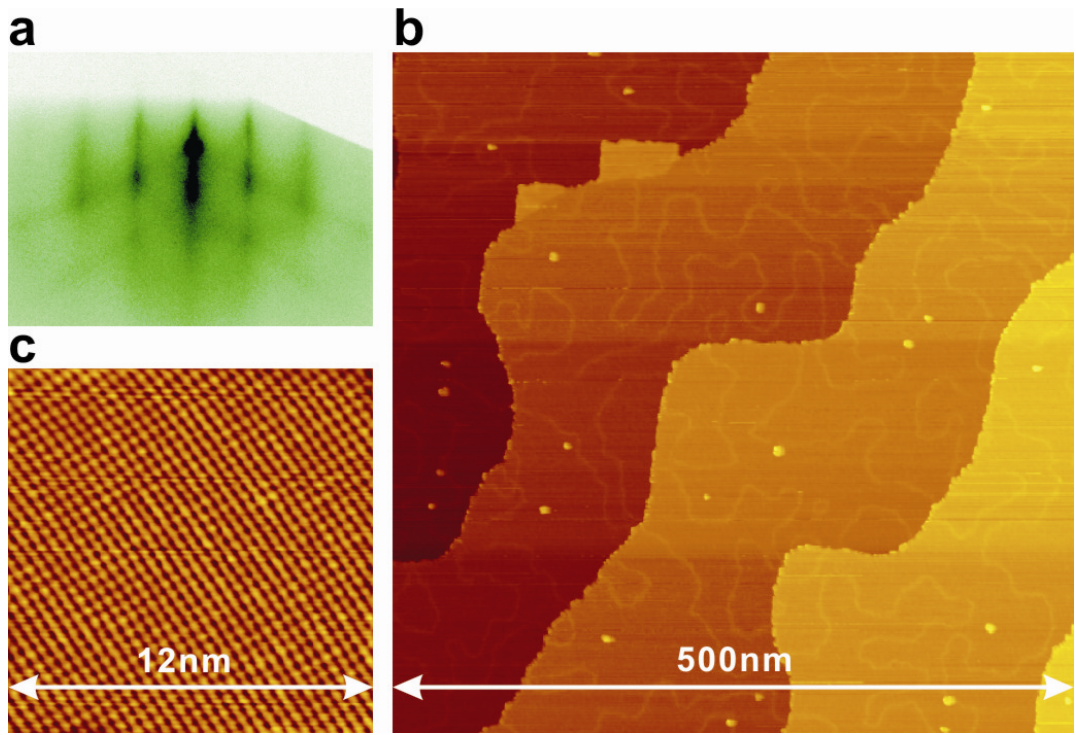

**Figure 1. Growth and characterization of high-quality single-layer film of FeSe. a**, RHEED pattern taken just after the growth of the film. **b,c**, Large- and atomic-resolved STM images of the film, taken at 7 K with $V_s$= 3.00 V and $I_t$ = 100 pA and at 3 K with $V_s$ = 0.05 V and $I_t$= 250 pA, respectively.

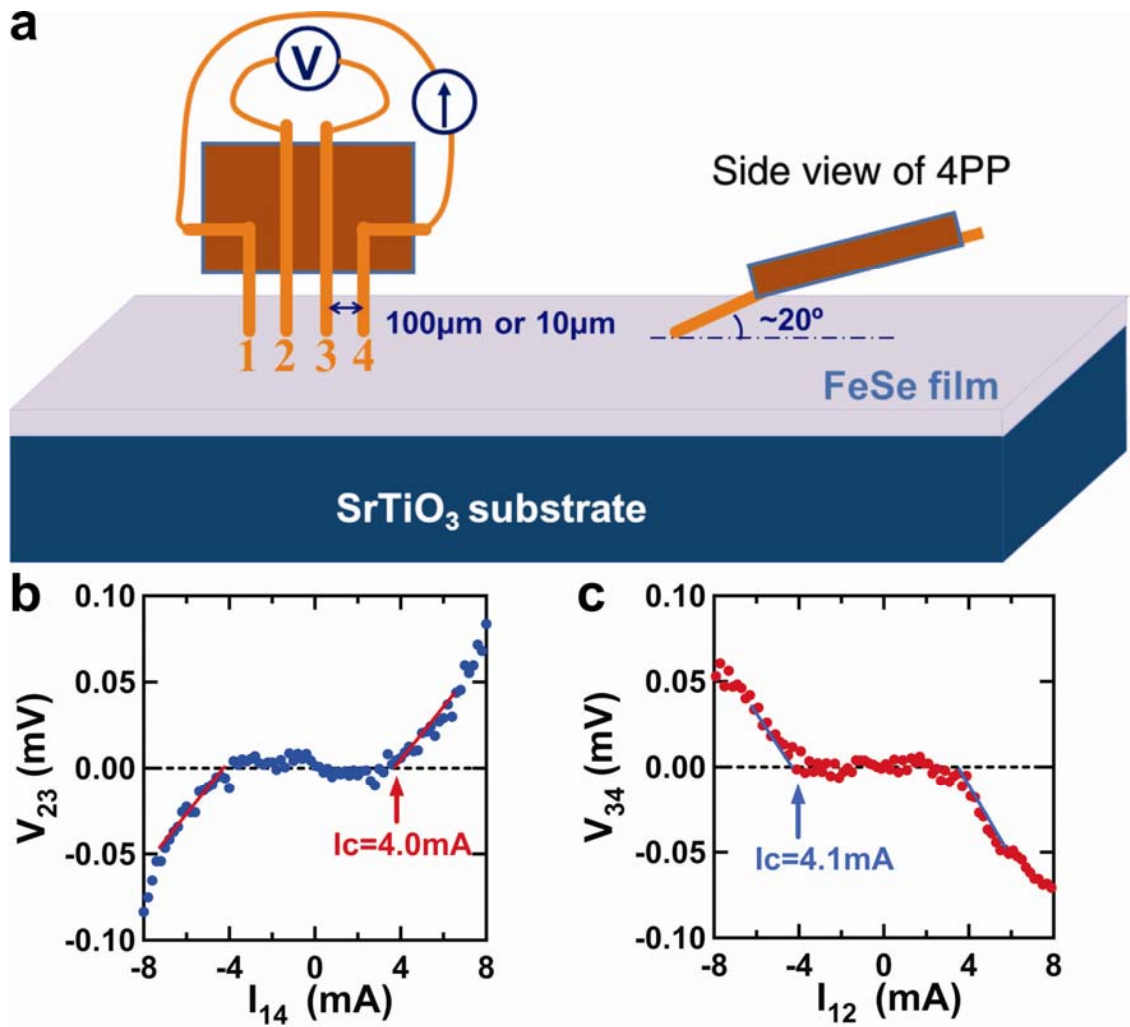

**Figure 2. a**, Schematic of 4PP transport measurement setup. The numbers are used to denote the contacting tips. All four tips can touch the sample surface gently in an inclined angle of ~20º. **b**,**c**, Typical superconducting I-V curves taken with a tip separation of 10 μm at 3.0 K with measurement configurations of C1423 and C1234, respectively.

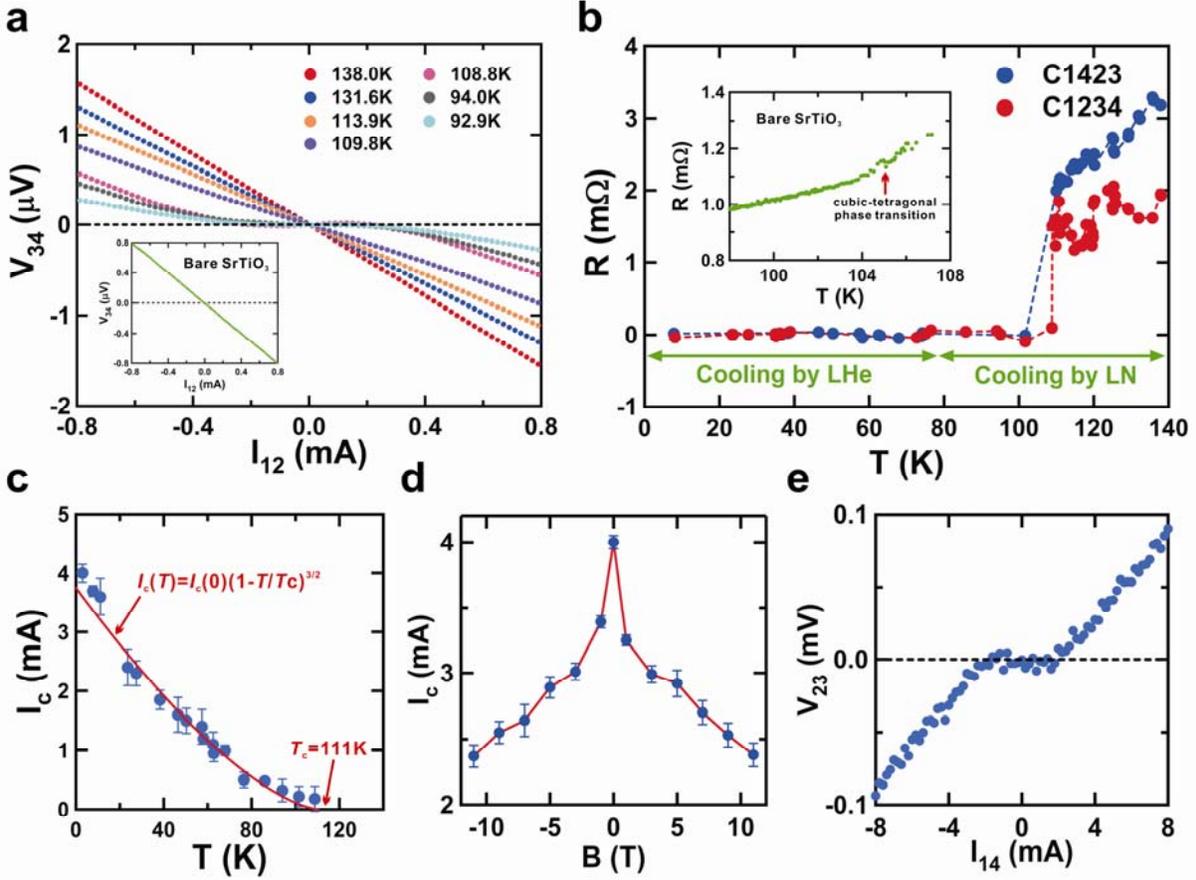

**Figure 3. Temperature and external magnetic field dependence of electrical transport property of a single-layer film of FeSe. a**, Typical I-V curves taken at various temperatures around 109 K. The inset shows a typical I-V curve of a bare STO surface taken at 99.2 K. **b**, Temperature dependence of the resistance obtained from a linear fit to the I-V curves. Above 79 K the sample was cooled by liquid liquid $N_2$. The inset shows the temperature dependence of resistance taken on a bare STO surface. **c**, Temperature dependence of $I_c$ retracted from the measured superconducting I-V curves. The superimposed curve is the fitted result following Ginsberg-Landau model. **d**, Out-of-plane magnetic field dependence of $I_c$ at 3 K. **e**, The I-V curve taken in a magnetic field of 11 T at 3 K. Data in **a-c** and **d-e** are taken with a 4PP with a 100 μm and 10 μm tip separation, respectively.

# Supplementary Information for "Superconductivity in single-layer films of FeSe with a transition temperature above 100 K"

Jian-Feng Ge, Zhi-Long Liu, Canhua Liu*, Chun-Lei Gao, Dong Qian, Qi-Kun Xue*, Ying Liu, Jin-Feng Jia*

[1]Key Laboratory of Artificial Structures and Quantum Control (Ministry of Education), Department of Physics and Astronomy, Shanghai Jiao Tong University, 800 Dongchuan Road, Shanghai 200240, China.

[2]Department of Physics, Tsinghua University, Beijing 100084, China.

[3]Department of Physics and Materials Research Institute, Pennsylvania State University, University Park, PA 16802, USA.

### Ohmic contact between 4PP and single-layer FeSe film

We reveal that all tips of the 4PP have Ohmic contacts with the single-layer FeSe films no matter a superconducting or a non-superconducting I-V curve is obtained. I-V curves were measured between each tip and the FeSe film after the tips touch the film one after another, as shown in Figs. S1 and S2. In both figures, part A shows the contact I-V curve measured when there is only one tip, which is Tip 4 in the present case, touched the FeSe film; part B and C show the contact I-V curves obtained when two and three tips have touched the FeSe film, respectively; part D was measured when all four tips have touched the FeSe film. Most of the above contact I-V curves are linear and give a resistance of several tens ohms. Sometimes we might get contact I-V curves showing a similar resistance but with very weak non-linear behavior. These curves would become linear when the tip was further pressed on the FeSe film, for example Tip 1 in Fig. S1 and Tip 2 in Fig. S2. Since the non-linear effect is very weak, it makes no influence on the 4PP transport measurement results.

### Superconducting and non-superconducting I-V curves measured with 4PP

In the experiment, both superconducting and non-superconducting I-V curves can be obtained in the 4PP transport measurement, as shown in Figs. S1g and S2e, respectively. At low temperature (below $T_c$), non-superconducting I-V curves emerge, possibly due to sample inhomogeneity or local damage of the sample around the tips. The latter case can be demonstrated by pressing the tips strongly on the FeSe film. While the contact I-V curves (Figs. S1d and S1e) make no intrinsic difference, the 4PP I-V curve changes from a superconducting one (Fig. S1g) to a non-superconducting one (Fig. S1h). However, the opposite change has never been observed in the experiment. For normal state characteristics, a certain resistance should result in a linear I-V curve which extrapolate to $V=0$ for $I=0$. Fig. S1f shows a large scale I-V curve taken at $T=90$ K, with dashed line fit the normal state nearly passing through (0, 0). In Fig. S1g (zoomed from Fig. S1f), I-V curve like those in Fig. 3a exhibits zero-resistance state more clearly. Because of the existence of contact resistance, we normally apply the current as low as possible to avoid heating effect.

## Control experiment on $Bi_2Sr_2CaCu_2O_{8+\delta}$

Optimally-doped $Bi_2Sr_2CaCu_2O_{8+\delta}$ (Bi2212) single crystal ($T_c \sim 90$ K) was used for transport experiment by the same 4PP probe in FeSe measurement. I-V curves were obtained at various temperatures. Figs. S3a and S3b show typical I-V curves at $T < T_c$ in C1423 and C1234, respectively. The critical current in both measurement configuration is estimated to be around 7.5 mA. These curves reveal a similar superconducting non-linear shape with FeSe/STO system.

## Difference between C1423 and C1234 measurement configurations

For a uniform infinite 2D film, the resistance measured in the two configurations is given by ref. 24

$$R_{1423} = \frac{V_{23}}{I_{14}} = \frac{R_s}{2\pi} \cdot \ln \frac{(r_3 - r_1)(r_4 - r_2)}{(r_4 - r_3)(r_2 - r_1)} \tag{SE1}$$

$$R_{1234} = \frac{|V_{34}|}{I_{12}} = \frac{R_s}{2\pi} \cdot \ln \frac{(r_3 - r_1)(r_4 - r_2)}{(r_3 - r_2)(r_4 - r_1)} \tag{SE2}$$

where $R_s$ is the sheet resistivity of the 2D film and $r_i$ is the position of the $i$th probe with the relation of $r_4 > r_3 > r_2 > r_1$. In case that the probe spacing is identical, i.e., $r_4-r_3=r_3-r_2=r_2-r_1$, the measured resistance can be simplified to be

$$R_{1423} = \frac{R_s}{2\pi} \cdot \ln 4 \tag{SE3}$$

$$R_{1234} = \frac{R_s}{2\pi} \cdot \ln \frac{4}{3} \tag{SE4}$$

For a uniform half-infinite 3D bulk material, the resistance measured in the two configurations is given by ref. 25

$$R_{1423} = \frac{V_{23}}{I_{14}} = \frac{\rho}{2\pi} \cdot \left( \frac{1}{r_2 - r_1} + \frac{1}{r_4 - r_3} - \frac{1}{r_3 - r_1} - \frac{1}{r_4 - r_2} \right) \tag{SE5}$$

$$R_{1234} = \frac{|V_{34}|}{I_{12}} = \frac{\rho}{2\pi} \cdot \left( \frac{1}{r_3 - r_2} + \frac{1}{r_4 - r_1} - \frac{1}{r_3 - r_1} - \frac{1}{r_4 - r_2} \right) \tag{SE6}$$

in which $\rho$ is the resistivity. In case that the probe spacing is identical, i.e., $r_4-r_3=r_3-r_2=r_2-r_1=s$, the measured resistance can be simplified to be

$$R_{1423} = \frac{\rho}{2\pi} \cdot \frac{1}{s} \tag{SE7}$$

$$R_{1234} = \frac{\rho}{2\pi} \cdot \frac{1}{3s} \tag{SE8}$$

It is seen from the above equations that when the probe spacing is not identical, $R_{1234}$ may be comparable to $R_{1423}$ if probe 2 and probe 3 have relatively smaller probe spacing in both 2D and 3D cases.

**Reproducibility of the superconducting I-V curves**

The superconducting I-V curves obtained on several samples prepared in the same way and cooled with liquid nitrogen are highly reproducible, although all data shown in Figs. 1-3 are from only one sample. Data in Figs. S1, S2 are from another FeSe/STO sample while those in S4 are from the third one. To have a detailed view of the resistance drop near $T_c$, we fixed the measuring location and took I-V curves while slowly raising the sample temperature. The typical I-V curves at various temperatures are shown in Fig. S4a. The R-T plot was obtained from a linear fit to the I-V curves and is shown in Fig. S4b, in which the sharp transition is very similar to that in Fig. 3b. The lower critical temperature is due to the sample quality although the growth process was the same. It is noted that the normal-state resistance shown in Fig. S4b is a little larger than that in Fig. 3b. This is attributed to the difference in the resistivity of the STO substrate, which is easily altered by annealing in sample preparation.

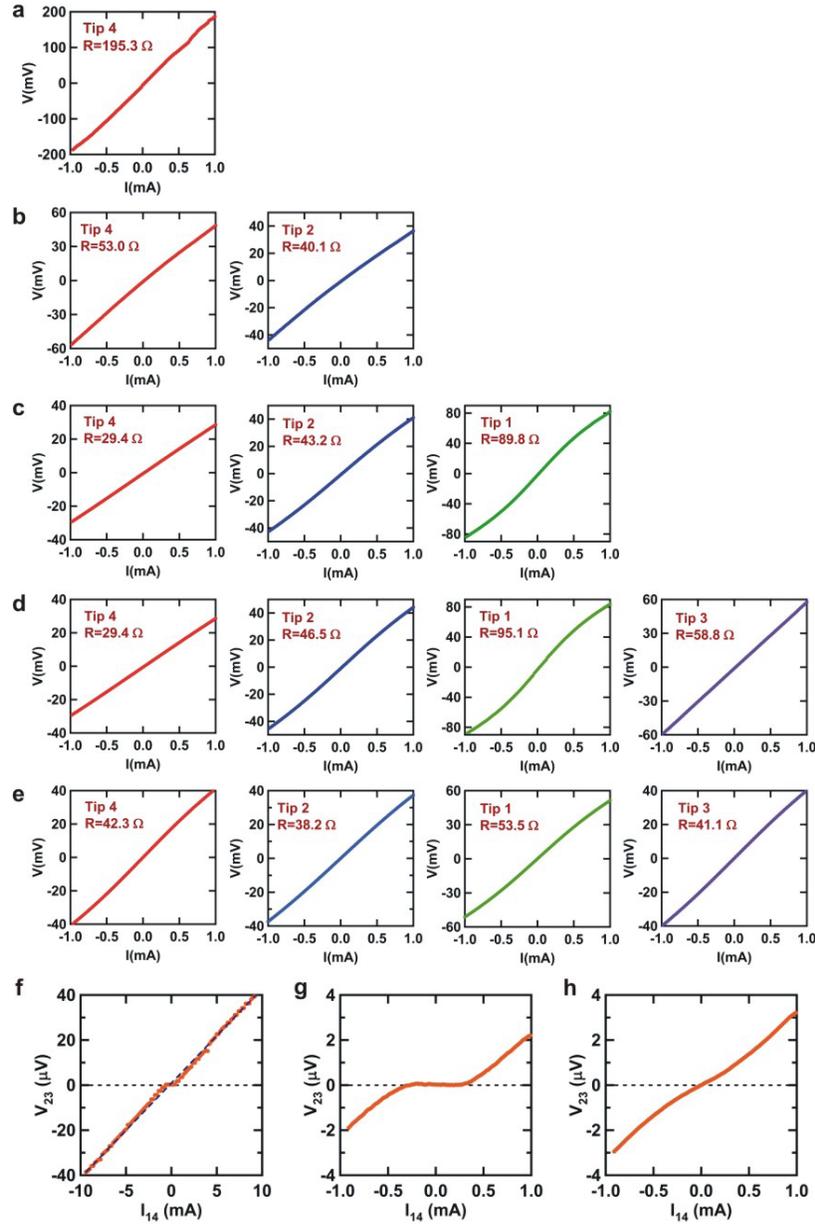

**Figure S1. a-d** Contact I-V curves between each tip and the single-layer FeSe film measured with the 100-μm-probe-separation 4PP at 90 K when one (**a**), two (**b**), three (**c**) and four (**d**) tips have touched the FeSe film. **e**, Contact I-V curves measured when the tips were strongly pressed on the FeSe film after the 4PP measurement of **f** and **g**. **f**, Typical large scale 4PP I-V curve showing superconducting behavior measured after the contact I-V measurement of **d**. **g**, Zoomed in view of **f**. **h**, 4PP I-V curve showing non-superconducting behavior measured after the contact I-V measurement of **e**.

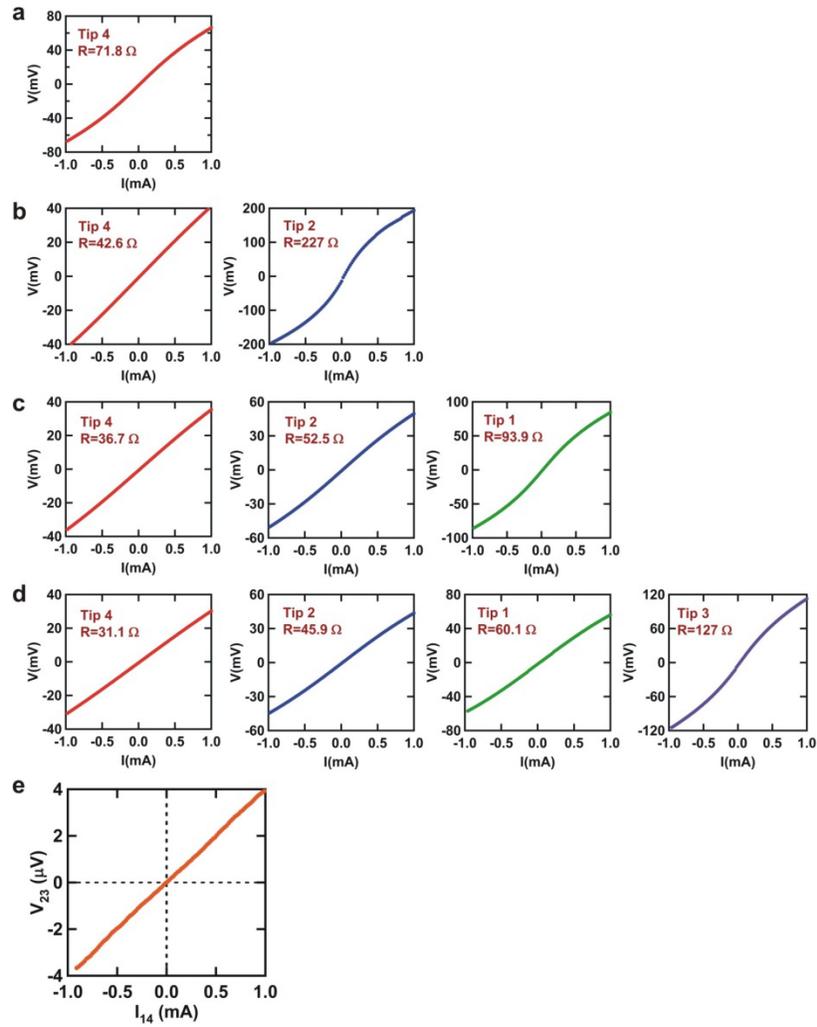

**Figure S2. a-d**, Contact I-V curves between each tip and the single-layer FeSe film measured by the 100-μm-probe-separation 4PP at 90 K when one (**a**), two (**b**), three (**c**) and four (**d**) tips have touched the FeSe film. **e**, The 4PP I-V curve showing non-superconducting behavior measured after the contact I-V measurement of **d**.

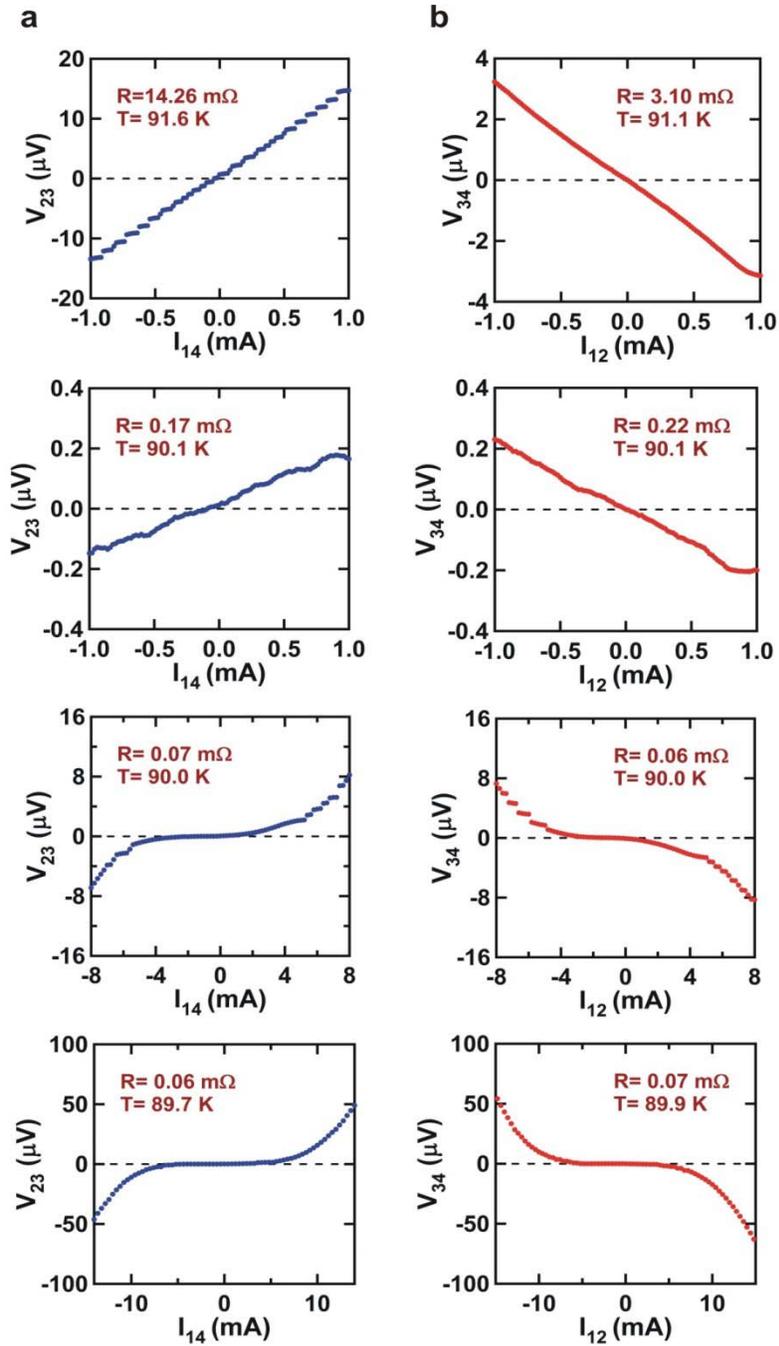

**Figure S3.** 4PP transport measurement on Bi2212 crystal with measurement configurations of C1423 (**a**) and C1234 (**b**), respectively. The sample temperatures and linearly fitted resistance around zero current are shown in each figure.

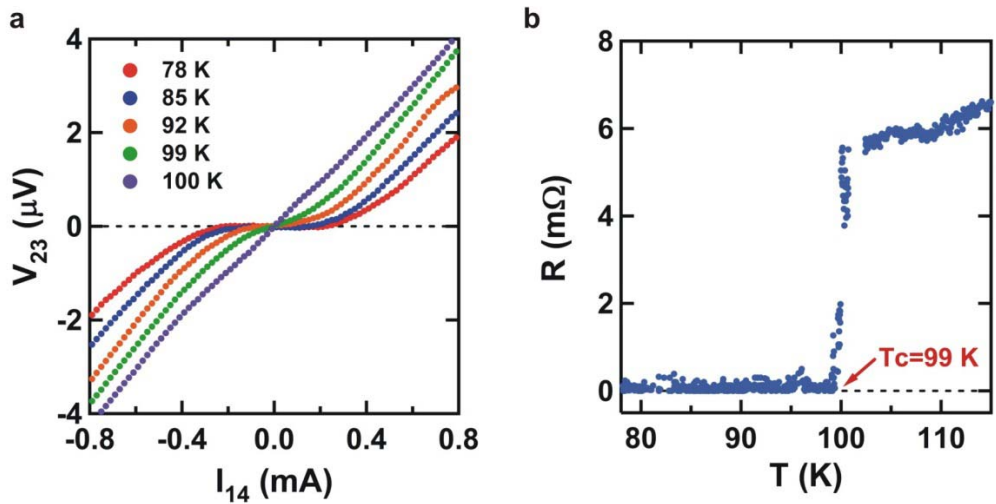

**Figure S4.** Temperature dependence of electrical transport property of another FeSe/STO sample measured with the 100-μm-tip-separation 4PP fixed at one measuring location while temperature was increasing. **a**, Typical I-V curves taken at various temperatures. **b**, Temperature dependence of the resistance obtained from a linear fit to the I-V curves. A critical temperature of 99 K was observed.